\documentclass[twocolumn]{aastex62}
\usepackage{CJK}
\usepackage{amsmath}
\usepackage{mathrsfs}

\newcommand{\ha}{\ifmmode {\rm H}\alpha \else H$\alpha$\fi}
\newcommand{\hb}{\ifmmode {\rm H}\beta \else H$\beta$\fi}
\newcommand{\oiii}{[\textrm{O}~\textsc{III}]}
\newcommand{\oiiilam}{[\textrm{O}~\textsc{III}]\ensuremath{\lambda5007}}

\newcommand{\nii}{[\textrm{N}~\textsc{II}]}
\newcommand{\niilam}{[\textrm{N}~\textsc{II}]\ensuremath{\lambda6583}}

\newcommand{\vasym}{v_{\text{asym}}}
\newcommand{\vasymbar}{\bar{v}_{\text{asym}}}
\newcommand{\re}{R_{\text{e}}}

\newcommand{\dSFR}{\text{SFR}_{\text{pair}}/\text{SFR}_{\text{CS}}}
\newcommand{\ssfr}{\text{sSFR}}

\newcommand{\kpc}{\ h^{-1}\ \text{kpc}}
\newcommand{\kms}{\ \text{km}\ \text{s}^{-1}}

\newcommand{\pdis}{d_{\text{p}}}
\newcommand{\dv}{\Delta v}

\received{July 20, 2019}
\revised{August 20, 2019}
\accepted{\today}
\submitjournal{ApJL}

\shorttitle{Kinematic Asymmetry of Galaxy Pairs}
\shortauthors{Feng et al.}

\begin{document}
\begin{CJK*}{UTF8}{gbsn}

\title{SDSS-IV MaNGA: Kinematic Asymmetry as An Indicator of Galaxy Interaction in Paired Galaxies}

\correspondingauthor{Shi-Yin Shen}
\email{ssy@shao.ac.cn}

\author[0000-0002-9767-9237]{Shuai Feng (冯帅)}
\affiliation{Key Laboratory for Research in Galaxies and Cosmology, Shanghai Astronomical Observatory, Chinese Academy of Sciences, \\
80 Nandan Road, Shanghai 200030, China}
\affiliation{University of the Chinese Academy of Sciences, No.19A Yuquan Road, Beijing 100049, China}

\author{Shi-Yin Shen (沈世银)}
\affiliation{Key Laboratory for Research in Galaxies and Cosmology, Shanghai Astronomical Observatory, Chinese Academy of Sciences, \\
80 Nandan Road, Shanghai 200030, China}
\affiliation{Key Lab for Astrophysics, Shanghai 200234, China}

\author{Fang-Ting Yuan (袁方婷)}
\affiliation{Key Laboratory for Research in Galaxies and Cosmology, Shanghai Astronomical Observatory, Chinese Academy of Sciences, \\
80 Nandan Road, Shanghai 200030, China}

\author{Rogemar A. Riffel}
\affiliation{Departamento de F\'isica, CCNE, Universidade Federal de Santa Maria, 97105-900, Santa Maria, RS, Brazil}
\affiliation{Laborat\'orio Interinstitucional de e-Astronomia - LIneA, Rua Gal. Jos\'e Cristino 77, Rio de Janeiro, RJ - 20921-400, Brazil}
\affiliation{Department of Physics \& Astronomy, Johns Hopkins University, Bloomberg Center, 3400 N. Charles St, Baltimore, MD 21218, USA}

\author{Kaike Pan}
\affiliation{Apache Point Observatory and New Mexico State University, P.O. Box 59, Sunspot, NM, 88349-0059, USA}

\begin{abstract}

The interaction between galaxies is believed to be the main origin of the peculiarities of galaxies, which disturbs not only the morphology but also their kinematics. These disturbed and asymmetric features are the indicators of galaxy interaction. We study the velocity field of the ionized gas of the paired galaxies in the SDSS-IV MaNGA IFU survey. Using the \texttt{kinemetry} package, we fit the velocity field of the ionized gas to quantify the degree of kinematic asymmetry. We find that the star formation rate (SFR) of the paired galaxies with high kinematic asymmetry is significantly enhanced even when the projected separation between the pair members is quite large ($\pdis \sim 100\kpc$). On the contrary, no significant SFR enhancement is found for the paired galaxies with low kinematic asymmetry even when their projected separation is small ($\pdis < 30 \kpc$). Moreover, we also find that the fraction of galaxies with high kinematic asymmetry is much higher in close pairs ($\pdis < 30 \kpc$) than those with larger $\pdis$, which explains well the early statistical finding of the significant SFR enhancement in close pairs. Our new findings illustrate that the kinematic asymmetry is an excellent indicator of galaxy-galaxy interaction strength, which helps us better understand the merging stage of the observed galaxy pairs.

\end{abstract}


\section{Introduction} \label{sec:intro}

In the hierarchical galaxy formation scenario, galaxies assemble most of their masses through galaxy-galaxy mergers. For major mergers, two comparable galaxies are first bounded by gravity and then form a galaxy pair. This pair status may last for a few Gyrs, where the strong galaxy-galaxy interaction significantly alters the physical properties of the member galaxies. In observation, galaxy pairs are usually selected with combined criteria, including the projected separation ($\pdis$) and line-of-sight velocity difference ($\dv$) \citep{Karachentsev1987,Ellison2008,Shen2016,Feng2019}. A large number of statistical studies have revealed that the galaxy pairs with small projected separations (e.g. $\pdis < 50 \kpc$) show significantly different features comparing with the field galaxies, such as the disturbed morphology \citep{Hernandez2005,Hernandez2006,Patton2016}, enhanced star formation rates (SFR) \citep{Ellison2008,Li2008,Patton2013}, diluted metallicities \citep{Kewley2006,MichelDansac2008,Scudder2012} and stronger nuclear activities \citep{Ellison2011,Liu2012,Satyapal2014}. All these observations suggest strong interactions between the pair members with small separations. Numerical simulations suggest that the peculiar physical properties of the paired galaxies are originated from the strong tidal effect during the pericenter passage between two galaxies \citep{Toomre1972,Mihos1996,Torrey2012,Moreno2015}.

Although the projected separation is a good indicator of the galaxy-galaxy interaction strength for a statistical sample, it may not be an ideal indicator of the galaxy-galaxy merging stage for individual galaxy pairs. First of all, the projected separation of two galaxies does not represent their physical separation because of the projection effect\citep{Soares2007}. On the other hand, pairs with the same physical separation do not necessarily have the same degree of interaction, which also depends on their merging stage (e.g., before or after their first passage) \citep{Torrey2012}. Therefore, to better characterize the interaction process between the merging galaxies, we need some other indicators, e.g., morphology \citep{Pan2019}. 

Numerical simulations show that, during two galaxies merging, their tidal force disturbs both their morphology and kinematic fields \citep{Hung2016}. The irregular kinematics of galaxies provides a clear signal of galaxy-galaxy interaction, which may happen even before the morphology disturbance. Indeed, recent integral field spectrograph (IFS) surveys such as CALIFA \citep{Sanchez2012} and SAMI \citep{Croom2012} have suggested that the galaxy-galaxy interaction is one of the main drivers of the irregular velocity field of galaxies \citep{Barrera2015,Bloom2017,Bloom2018}. Therefore, we expect that the kinematic field of the paired galaxies would be a good indicator of the galaxy-galaxy interaction as well as the projected separation. By including the parameters of kinematic fields, the details of the galaxy merging process may be better depicted. 

In this work, we study the kinematic asymmetry of the ionized gas of a large sample of paired galaxies using the MaNGA (Mapping Nearby Galaxies at APO) data \citep{bundy2015,law2015,wake2017} in the Sloan Digital Sky Survey IV \citep[SDSS-IV][]{blanton2017,smee2013,Gunn2006}. The paper is organized as follows. In Section \ref{sec:data}, we first introduce the galaxy pair sample and the measurement of kinematic asymmetry using MaNGA data. Next, we show the results of the data analysis in Section \ref{sec:result}, and make relevant discussions in Section \ref{sec:dis}. Finally, a brief summary is listed in Section \ref{sec:sum}. Throughout this paper, we adopt a standard cosmology, with $\Omega_M=0.3$, $\Omega_\Lambda=0.7$, and $\text{H}_0=100~h~\kms~\text{Mpc}^{-1}$ with $h=1$.

\section{Data} \label{sec:data}

\subsection{Paired Galaxy Sample}

The galaxy pair sample are taken from \citet{Feng2019}, which selects isolated galaxy pairs in the main sample galaxies of SDSS with the following criteria: (1) the projected separation: $10 \kpc \leq \pdis \leq 200 \kpc$; (2) the line-of-sight velocity difference: $|\Delta V| \leq 500 \kms$; (3) each pair member only has one neighbor satisfying above criteria. 

Among the galaxy pair sample, $1398$ member galaxies have been observed and processed by MaNGA Product Launch 8 (MPL-8) \citep{drory2015,law2016,yan2016a,yan2016b}. In this study, we only consider star-forming galaxies, whose velocity and star formation rate maps could be well quantified from the IFS data.  We use a criterion, the global specific star formation rate $\log(\ssfr/\text{yr}^{-1})>-11$ (taken from the GALEX-SDSS-WISE Legacy Catalog 2  \citep{salim2016,salim2018} ) to select the star-forming galaxies and have obtained 632 of them. We take the advanced products of the MaNGA Data Analysis Pipeline \citep[DAP,][]{westfall2019} to obtain the $\ha$ velocity map and $\ha,\hb,\oiiilam,\niilam$ flux maps \footnote{The emission line fluxes are measured from Gaussian fit \citep{Belfiore2019}} for each galaxy.

\subsection{Measurement of Kinematic Asymmetry}\label{sec:kin}

We use the \texttt{kinemetry} package\footnote{\url{http://davor.krajnovic.org/idl/}} to fit the H$\alpha$ velocity map for each galaxy in our sample \citep{kinemetry}. This package divides the velocity map into a sequence of concentric elliptical rings according to the pre-defined parameters, including galaxy center, kinematic position angle, and ellipticity. Then, it fits the velocity values in each ellipse to the Fourier series:
\begin{equation}
    V(a,\psi)=A_0(a)+\sum^N_{n=1}k_n(a)\cos[n(\psi-\phi_n(a))]\,,
\end{equation}
where $\psi$ is the azimuthal angle in the galaxy plane, $a$ is the semi-major axis of the ellipse, and $A_0$ is zero-order Fourier component. The parameters $k_n$ and $\phi_n$ are the amplitude and the phase coefficient of $n$th order Fourier component, respectively. The first-order coefficient $k_1$ describes the symmetric pattern of the velocity map, which is typically contributed by the rotating motion of galaxies, while high-order coefficients describe the asymmetric pattern of the velocity map, indicating the contribution from non-rotating motion. Thus, the kinematic asymmetry at a given radius is expressed as 
\begin{equation}
    \vasym=\frac{k_{2}+k_{3}+k_{4}+k_{5}}{4k_{1}}\,.
\end{equation}
The characteristic value of the kinematic asymmetry for the entire galaxy, denoted as $\vasymbar$, is represented by its average value within 1 $\re$. With this definition, the larger value of $\vasymbar$ means higher asymmetry of the velocity field. A galaxy with a higher $\vasymbar$ value means that the contribution of non-rotating motion to the velocity maps is higher. 

In practice, the kinematic position angle and ellipticity of concentric elliptical rings are fixed as the best fit values from the single Sersic fit of $r$-band photometric image ($\texttt{SERSIC\_PHI}$ and $\texttt{SERSIC\_BA}$ from NSA catalog \footnote{\url{http://www.nsatlas.org/}}). We take the center of the MaNGA velocity map as the center of kinemetry fitting. Few galaxies whose center of velocity map is not located at the photometric center are not taken into account. We only fit the galaxies, which have more than $70\%$ spaxels with the S/N of H$\alpha > 5$ in 1.5$\re$. Finally, $578$ paired galaxies are kept in our sample.

\subsection{Sub-samples and Control Sample}

\begin{figure*}
\includegraphics[width=\linewidth]{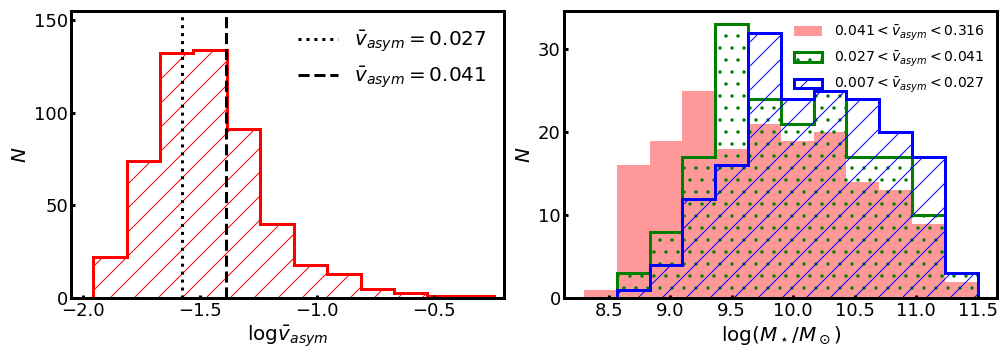}
\caption{Left panel: The $\vasymbar$ distribution of 578 star-forming paired  galaxies in MaNGA MPL-8. The black dashed and dotted lines indicate the $\vasymbar=0.027$ and $\vasymbar=0.041$, respectively. Right panel: The stellar mass distributions of three sub-samples defined by $\vasymbar$ values.} 
\label{fig:vasym_pdf}
\end{figure*}

We show the $\vasymbar$ distribution of the 578 paired galaxies with the hatched histogram in the left panel of Figure \ref{fig:vasym_pdf}, and separate these paired galaxies into three equal size sub-samples according to their $\vasymbar$ values. The $\vasymbar$ intervals of three sub-samples are $0.007 < \vasymbar < 0.027$ (low asymmetry, LA), $0.027 < \vasymbar < 0.041$ (medium asymmetry, MA) and $0.041 < \vasymbar < 0.316$ (high asymmetry, HA), respectively. The dotted and dashed vertical lines represent two thresholds ($\vasymbar=0.027$ and $\vasymbar=0.041$) respectively.  We also show the distribution of stellar mass of three sub-samples in the right panel of Figure \ref{fig:vasym_pdf}. The blue, green and red histograms represent the LA, MA and HA respectively. It is clear that the fraction of lower mass galaxies of HA is higher than MA and LA, which is consistent with the finding of \citet{Bloom2017} for the general galaxy population.

The control sample of galaxies are selected from the non-paired star-forming galaxies ($\log(\ssfr/\text{yr}^{-1})>-11$) by matching their stellar mass and redshift to the paired galaxies one-by-one with $|\Delta \log (M_\star/M_\odot)| < 0.2 $ and redshift $|\Delta z| < 0.01$. Specifically, the non-paired galaxies are also selected from MaNGA MPL-8 and are defined as those without bright neighbors ($r<17.77$) within the interval of $\pdis \leq 200 \kpc$ and $|\Delta V| \leq 500 \kms$ \citep{Feng2019}. Since the galaxies with $\vasymbar$ measurement might be biased towards these objects with strong emission lines, we also require the control galaxies to have more than 70\% spaxels with 5-$\sigma$ detection of $\ha$ flux in $1.5\re$ as that done for the paired galaxies in Section \ref{sec:kin}. 

\section{Properties of Paired Galaxies with Different Kinematic Asymmetry} \label{sec:result}

In this section, we compare the physical properties of three sub-samples of paired galaxies and explore their connection with kinematic asymmetry. 

\begin{figure*}
\includegraphics[width=\linewidth]{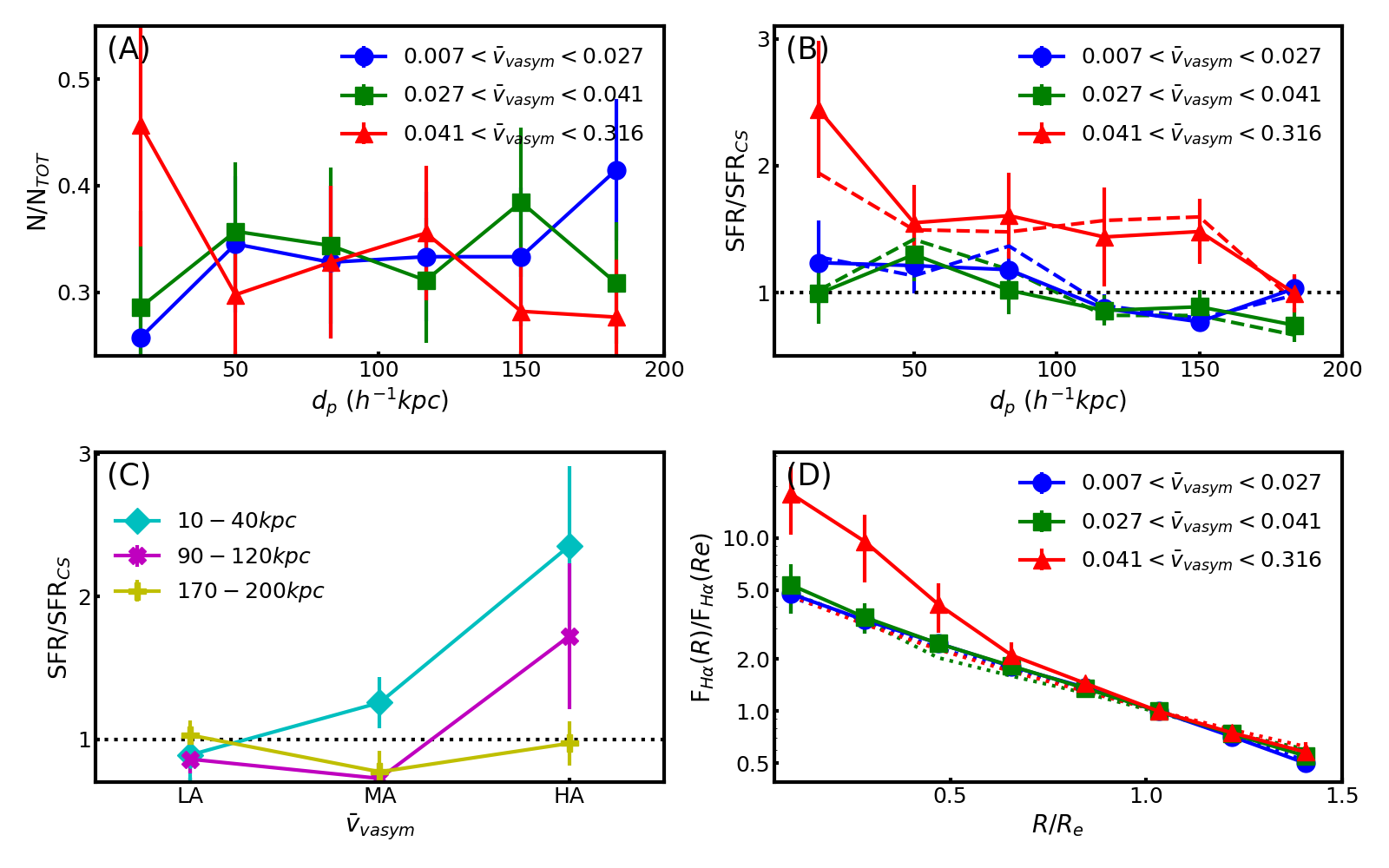}
\caption{(A): Fraction of galaxies in each sub-sample as a function of $\pdis$. (B): The strength of SFR enhancement as a function of $\pdis$. The solid lines show the results of all paired galaxies, while the dashed lines represent the paired galaxies without bars. (C): The strength of SFR enhancement as a function of $\vasymbar$. (D): Radial profile of H$\alpha$ flux of paired galaxies. The dotted lines represent control samples. In (A) (B) and (D), the three color-coded lines indicate the three sub-samples classified by $\vasymbar$ value, where the symbols represent the mean values and the error bars indicate the uncertainties of the mean values obtained from bootstrap sampling. In (C), the three color-coded lines represent three $\pdis$ bins. }
\label{fig:allpair}
\end{figure*}


\subsection{Projected Separation}\label{sec:pdis}

We first compare the fraction of three $\vasymbar$ sub-samples in different $\pdis$ bins and show the result in the panel (A) of Figure \ref{fig:allpair}. The solid blue circles, green squares, and red triangles represent the sub-samples of LA, MA, and HA, respectively. The error bars are estimated from the bootstrap sampling of the paired galaxy sample. In the largest $\pdis$ bin ($\pdis \sim 200\kpc$), the fraction of LA is the largest (larger than $40\%$), while the fraction of HA is the lowest (lower than $30\%$). For the smallest $\pdis$ bin ($\pdis < 30\kpc$), the fraction of HA becomes the largest, which increases up to $45\%$. At the same time, the fractions of MA and LA drop to $25\%$. We mention that the stellar mass distribution of our paired galaxies is almost independent of $\pdis$. Therefore, this result is not influenced by the different distributions of stellar mass of three sub-samples.

These results show that the fraction of paired galaxies with high kinematic asymmetry is correlated with projected separation. Paired galaxies with smaller projected separation are more likely to have a highly asymmetric kinematic field. Especially for $\pdis < 30 \kpc$, the fraction of paired galaxies with a highly asymmetric kinematic field is close to half. 

The higher fraction of HA galaxies in closer pairs is normally attributed to the stronger tidal force from companion galaxies \citep{Barrera2015,Bloom2018}. Statistically, paired galaxies with smaller physical separation would also show smaller projected separation. In the simplest case, the tidal force is inversely proportional to the cube of the physical separation between the paired members. As the decrease of the separation, the tidal force increases dramatically. However, this simple scenario can not give a full explanation of the observed trends we have discussed.

During the galaxy merging process, the physical separation decreases until the pericenter passage and then increases until reaching the apocenter \citep{Torrey2012}. On the other hand, the disturbance of the velocity field not only depends on the tidal force (acceleration) but also on the interaction time. As a result, because of longer interaction time, paired galaxy after the pericenter passage would show a more disturbed velocity field than that before the passage \citep{Hung2016}. Also, because of the energy dissipation from dynamical friction, paired galaxies after the pericenter passage, on average, have a smaller physical separation than those before the passage. Putting all these effects together, we see that paired galaxies with smaller $\pdis$ show a higher fraction of HA galaxies. This scenario also explains that there are about half of the close pairs ($\pdis < 30 \kpc$) do not show high kinematic asymmetry. These LA/MA galaxies in close pairs are either caused by projection effect or before the pericenter passage. On the other hand, a significant fraction of the large separation pairs ($\pdis > 100\kpc$) shows high kinematic asymmetry. These HA galaxies are possibly at the stage after the pericenter passage and may approach the apocenter, resulting in a disturbed velocity field and a large separation with the companion galaxy (see more discussion in Section \ref{sec:totsfr}).


\subsection{Total SFR}\label{sec:totsfr}

In this section, we explore the correlation between the kinematic asymmetry and the enhancement of star formation in paired galaxies. We use the ratio of the total SFRs of the paired galaxies to their corresponding control galaxies, $\dSFR$, to represent the SFR enhancement in paired galaxies. 

The panel (B) of Figure \ref{fig:allpair} displays the SFR enhancement of three sub-samples with different kinematic asymmetry as a function of $\pdis$. Generally, HA galaxies show significant SFR enhancement at $\pdis<150\kpc$. The $\dSFR$ reaches about $250\%$ in the smallest $\pdis$ bin ($\pdis<30\kpc$). At very large $\pdis$ ($\pdis>100\kpc$), the $\dSFR$ is still at the level of $150\%$. In contrast, the enhancement of the total SFRs of the other two sub-samples (MA and LA) is not significant, even in the smallest $\pdis$ bin ($\pdis<30\kpc$). 

We also show the SFR enhancement as a function of kinematic asymmetry within given $\pdis$ intervals in the panel (C). The $\vasymbar$ bins follow the intervals of three sub-samples. For clarity, we show three cases of $\pdis$ intervals: $10\kpc<\pdis<40\kpc$, $90\kpc<\pdis<120\kpc$, $170<\pdis<200\kpc$, which represent the galaxy pairs with very small, medium and large projected separation respectively. In the largest $\pdis$ interval, we see there is no SFR enhancement regardless of $\vasymbar$ values. This result is in good consistency with the recent study of \citet{Feng2019}, where the member galaxies in pairs with $\pdis\sim 200\kpc$ are shown with few interactions through a strictly statistical approach. Nevertheless, there is still a small but significant fraction ($\sim 30\%$) of HA galaxies at such large $\pdis$ in the panel (A) of Figure \ref{fig:allpair}, which don't show SFR enhancement at all. The HA features of these galaxies might be caused by the internal process (e.g., bar effect, see more discussion in Section \ref{sec:dis}) and/or the peculiarities of the galaxies themselves instead of galaxy interactions. However, a detailed study of the HA features of these galaxies is out the scope of this study. For these paired galaxies within intermediate $\pdis$ interval, the majority of them (LA and MA, $\sim 65\% $) still show no enhanced star formation, which is also consistent with early finding that the paired galaxies out to $\pdis > 80 \kpc$ in general show very weak enhanced star formation \citep{Scudder2012,Patton2013}. However, in this case, there are a fraction of galaxies (HA, $35\%$) that indeed show significantly enhanced star formation ($\dSFR \sim 150\%$). As we have discussed, these HA galaxies might be in the stage of being after the first pericenter passage and approaching the apocenter. Among the galaxies with very close companions ($\pdis<40\kpc$), the LA galaxies do not show enhanced star formation at all. While the MA galaxies show moderate SFR enhancement ($\dSFR \sim 120\%$), and the HA galaxies show the highest $\dSFR$ ($\sim 250\%$). According to the discussion in Section \ref{sec:pdis}, the LA galaxies might be affected by the projection effect, i.e., there are not in real close galaxies pairs. On the other hand, although both the MA and HA galaxies are in real pairs, they might still be affected by the projection effect differently or at the different stages of the merging process, which are before and after the first passage, respectively. 

\begin{figure}
    \centering
    \includegraphics[width=\columnwidth]{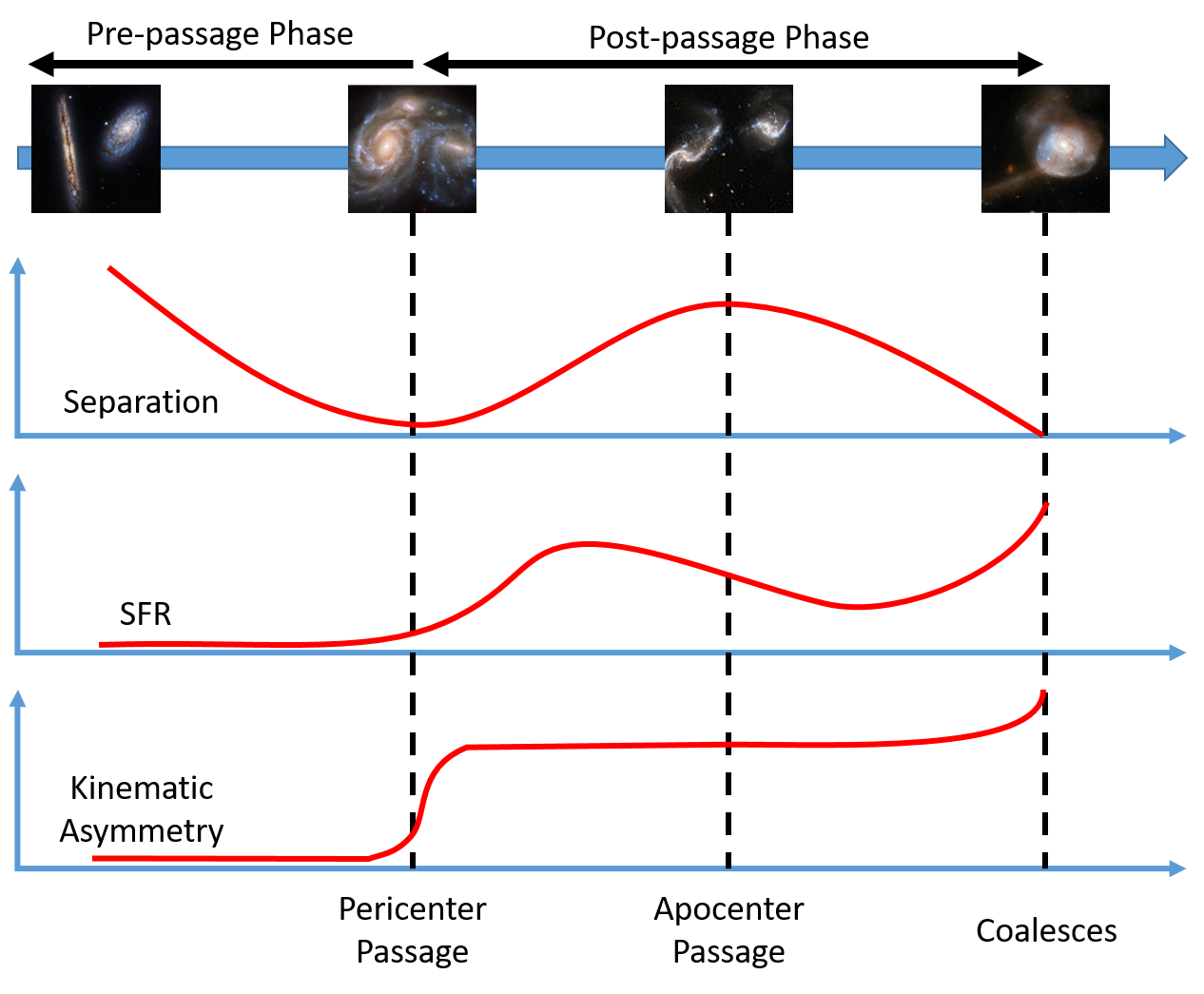}
    \caption{Illustration of the merging stage of galaxy pairs. In this scenario, before the final coalescence, a galaxy pair will experience the pre-passage stage and post-passage stage, which is separated by the pericenter passage. The red solid curves indicate the evolutionary trends of physical separation, total SFR, and kinematic asymmetry of the pair members during merging. The black dashed lines represent the occurrence time of pericenter passage, apocenter passage, and coalescence, respectively. }
    \label{fig:Carton}
\end{figure}

The SFR enhancement results shown in Panel (B), and (C) are in good agreement with the merging stage scenario we discussed in Section \ref{sec:pdis}. Numerical simulations suggest that the SFR enhancement mainly happens after the pericenter passage of galaxy merging \citep{Moreno2015}. It means, only at the late stage of galaxy merging (e.g., after the first pericenter passage), there is enough time to disturb both the ionized gas (to induce kinematic asymmetry) and neutral gas (to enhance star formation).  To sum up, we outline the global galaxy merging process below, and illustrate it with a schematic diagram in Figure \ref{fig:Carton}.

During the hierarchical structure formation, two galaxies begin to form a galaxy pair and have interactions on each other, starting from a projected separation out to $150\kpc$. In the process of galaxies approaching each other, the tidal force increases, and the velocity field starts to be disturbed. Then, at the stage of the first pericenter passage, the tidal force reaches the maximum, and the member galaxies show moderate velocity asymmetry and enhanced star formation.  After the first pericenter passage, accompanying with the gas consumption, the tidal force decreases and so that the enhanced star formation might also gradually decrease, while the disturbed velocity field (morphology) is kept. At the very late stage of galaxy merging, i.e., at the second pericenter passage or right before the final coalescence, these paired galaxies have very small separation (projected separation), suffer the strongest tidal effect, and therefore show the highest SFR enhancement. 

Within this merging stage scenario,  the statistical correlation between the SFR enhancement and the projected separation reported in many previous works \citep[e.g.][]{Scudder2012,Patton2013} is a natural conclusion. Moreover, only the galaxies with both small projected separation and high velocity asymmetry are in the late stage of the galaxy merging and show the strongest enhanced star formation. As a corroboration, the average 
$\dSFR $ is about $140\%$ for the galaxies in close pairs in general \citep{Ellison2008,Feng2019}, while  $\dSFR $ is as high as $250\%$ for these HA galaxies in close pairs.


\subsection{Radial Profiles of SFR}

In Section \ref{sec:totsfr}, we have shown that only the paired galaxies with high kinematic asymmetry show significant SFR enhancement. In this section, we take advantage of IFS data to explore further where the enhanced SFR happens.

We use the radial profiles of $\ha$ flux to represent the SFR profiles. For each galaxy, we take $\ha$ flux maps from MaNGA DAP and bin it into a sequence of concentric elliptical rings. The position angle and ellipticity of concentric elliptical rings are taken from the $\texttt{SERSIC\_PHI}$ and $\texttt{SERSIC\_BA}$ of $r$-band in NSA catalog. The semi-major axes of elliptical rings span from $0.1\re$ to $1.5\re$, where  $\re$ is the effective radius of $r$-band image for the single Sersic fitting. We only use the spaxels with reliable emission line flux measurements (S/N $>5$ for $\ha$, and S/N$>3$ for $\hb$, $\oiii$ and $\nii$) and that are classified as star-forming regions according to the criteria of \citet{Kauffmann2003} in BPT diagram. We make corrections on $\ha$ flux for each star-forming spaxel using the Balmer decrement: 
\begin{equation}
A(\ha)= 6.56\log_{10}[\frac{(\ha/\hb)_{\text{obs}}}{2.86}],
\end{equation}
where the intrinsic line ratio of $\ha/\hb$  is assumed to be 2.86, and the attenuation curve is adopted as \citet{Calzetti2000}. Then, the radial profile of $\ha$ flux of each paired galaxy, $F_{\ha}(R)$ is obtained by calculating the mean values of the star-forming spaxels enclosed by the elliptical rings. To simplify the comparison in the next steps, we normalize the $\ha$ flux profile of each galaxy with its $\ha$ flux at an effective radius $F_{\ha}(\re) $. Finally, we take the mean values of  $F_{\ha}(R)/F_{\ha}(\re)$  of a sub-sample of galaxies to represent their average SFR profiles.

We show the relative SFR profiles of three sub-samples of paired galaxies in the panel (D) of Figure \ref{fig:allpair}. The SFR profiles of their corresponding control galaxies are also plotted as dotted lines for comparison. The three samples of control galaxies, although with different stellar mass distributions (right panel of Figure \ref{fig:vasym_pdf}), show almost identical SFR profiles after scaled with their effective radii $\re$.  Also, as expected, the SFR profiles of the LA and MA galaxies are very similar to the control galaxies because of their negligible SFR enhancement.  For the HA galaxies, the radial profile of the outer region ($R>0.8\re$) is almost the same as the control galaxies,  while it is significantly enhanced in the inner region ($R<0.8\re$). Combining the enhancement of total SFR shown in the panel (B) and (C) of Figure \ref{fig:allpair} (Section \ref{sec:totsfr}), we further conclude that the SFR enhancement of HA galaxies mainly happens in their inner region. Our finding is consistent with \citet{Moreno2015}, where the enhanced SFR in paired galaxies is found inside a few kpcs (see also \citealt{Pan2019}). 

The inner SFR enhancement ($R<0.8\re$) shown in panel (D) is very likely due to the increase of the gas density caused by the tidal-induced gas inflow. In the paired galaxies, the tidal-induced gas inflow usually happens after the pericenter passage \citep{Torrey2012}, which then contributes an asymmetric component into its velocity field, and so that increases the $\vasymbar$ value. Not only that, the inflowing gas also increases the gas density of the inner regions \citep{Barnes1996,Moreno2015} and then enhance the star formation there in short time-scales \citep{Feng2019}. Therefore, this radial dependent SFR enhancement result reinforces the merging stage scenario we proposed in Section \ref{sec:totsfr}. 

\section{Discussion: Bar-induced Asymmetry}\label{sec:dis}

Kinematic asymmetry is mainly contributed by non-rotating motion. Besides the tidal-induced gas disturbance, the bar-driven gas inflow also enhances non-rotating components of the velocity field \citep{Regan1999}. Moreover, many observations have suggested that barred galaxies also show enhanced SFR \citep{Chown2019}. To test whether the barred galaxies play a role in our study, we further check the correlation between the velocity field asymmetry and enhanced SFR for the non-bar paired galaxies. We take the bar and non-bar classification from the Galaxy Zoo project \citep{Willett2013,Hart2016}. In our 578 paired galaxies, 270 galaxies have reliable classification. Among them, 227 are non-bar, and 53 are barred galaxies.

We repeat all the earlier analysis of these non-bar galaxies. The results for non-bar galaxies are almost the same as the results for all paired galaxy sample. As an example, we show the enhanced SFR as a function of the projected distance for non-bar galaxies with dashed lines in panel (B) of Figure \ref{fig:allpair}. Therefore, we conclude that the increase of $\vasymbar$ and SFR in paired galaxies are mainly caused by the tidal-induced gas inflow rather than the bar effect.  Does the bar-phenomena play any roles in galaxy pairs? Taking a preliminary look, the fraction of barred galaxy in our paired galaxy sample is $19 \pm 3\%$, which is slightly larger than the control sample ($15 \pm 1\%$). That indicates interactions between pair members may induce the bar structure \citep{Nicolas2019}. The detailed answer to this question, however, is beyond the scope of this study.

\section{Summary} \label{sec:sum}

We select $632$ paired star-forming galaxies from the MaNGA survey. Using \texttt{kinemetry} package, we successfully fit the $\ha$ velocity map of 578 galaxies and quantify their kinematic asymmetry by the parameter $\vasymbar$. Then, we separate these galaxies into three sub-samples according to the $\vasymbar$ and compare their physical properties.

First, We find that the fraction of galaxies with large $\vasymbar$ values is much higher in close pairs ($\pdis < 50\kpc$) than in pairs with larger separations. Second, for the total SFR, only the paired galaxies with large $\vasymbar$ values have significant enhancement comparing to isolated galaxies. In contrast, there is little SFR enhancement in the paired galaxies with small $\vasymbar$ values, even for $\pdis < 50\kpc$. Third, the SFR enhancement of paired galaxies with large $\vasymbar$ values mainly happens in the inner region of galaxies ($R<0.8\re$). 

From these results, we suggest that the kinematic asymmetry is a better indicator of galaxy-galaxy interaction than the projected separation, which is commonly used in statistical studies of galaxy pairs. The paired galaxies with high kinematic asymmetry are very likely at the stage after pericenter passage. During this stage, the tidal-induced inflow significantly increases the gas density at the inner region of galaxies and enhances the star formation there in short time scales.

In our scenario, tidal effects first produce the acceleration, change the velocity field, and then disturb the morphology. Therefore, the disturbed morphology is also a good indicator of the interactions between galaxy pairs \citep{Barrera2015,Pan2019}. Correlations between the kinematic asymmetry and disturbed morphology have also been found \citep{Hung2016,Bloom2017}. Comparing with kinematics measurement, photometric morphology is much cheaper. Nevertheless, we emphasize the distortion of the velocity field happens on shorter time scales than the distortion of morphology. Also, given the accuracy of the velocity measurement is up to a few kilometers per second for typical resolution galaxy spectroscopy, we conclude that the velocity field asymmetry measurement is a more sensitive indicator of galaxy interaction than morphology. 

In the future, by combining the analysis of the velocity field and morphology of the paired galaxies together and comparing it with the numerical simulations, it is possible to establish a complete and detailed merging scenario of galaxies.

\acknowledgments

This work is partly supported by the National Natural Science Foundation of China (NSFC) under grant nos.11433003, 11573050, and by a China-Chile joint grant from CASSACA. SY thanks support from National Key R\&D Program of China No.2019YFA0405501. FTY acknowledges support from the Natural Science Foundation of Shanghai (Project Number: 17ZR1435900). 

Funding for the Sloan Digital Sky Survey IV has been provided by the Alfred P. Sloan Foundation, the U.S. Department of Energy Office of Science, and the Participating Institutions. SDSS-IV acknowledges support and resources from the Center for High-Performance Computing at the University of Utah. The SDSS web site is \url{www.sdss.org}.

SDSS-IV is managed by the Astrophysical Research Consortium for the Participating Institutions of the SDSS Collaboration including the Brazilian Participation Group, the Carnegie Institution for Science, Carnegie Mellon University, the Chilean Participation Group, the French Participation Group, Harvard-Smithsonian Center for Astrophysics, Instituto de Astrof\'isica de Canarias, The Johns Hopkins University, Kavli Institute for the Physics and Mathematics of the Universe (IPMU) / University of Tokyo, the Korean Participation Group, Lawrence Berkeley National Laboratory, Leibniz Institut f\"ur Astrophysik Potsdam (AIP),  Max-Planck-Institut f\"ur Astronomie (MPIA Heidelberg), Max-Planck-Institut f\"ur Astrophysik (MPA Garching), Max-Planck-Institut f\"ur Extraterrestrische Physik (MPE), National Astronomical Observatories of China, New Mexico State University, New York University, University of Notre Dame, Observat\'ario Nacional / MCTI, The Ohio State University, Pennsylvania State University, Shanghai Astronomical Observatory, United Kingdom Participation Group, Universidad Nacional Aut\'onoma de M\'exico, University of Arizona, University of Colorado Boulder, University of Oxford, University of Portsmouth, University of Utah, University of Virginia, University of Washington, University of Wisconsin, Vanderbilt University, and Yale University.

\bibliographystyle{aasjournal}
\bibliography{ref.bib}

\end{CJK*}
\end{document}